\documentclass[12pt]{article}

\usepackage{amsmath,amssymb}
\usepackage{graphicx}

\newcommand{\Tr}{\operatorname{Tr}} 
\newcommand{\smallfrac}[2]{\mbox{\small ${\displaystyle \frac{#1}{#2}}$}}

\long\def\symbolfootnote[#1]#2{\begingroup%
\def\thefootnote{\fnsymbol{footnote}}\footnote[#1]{#2}\endgroup}
\renewcommand{\thefootnote}{\dag}
\begin{document}
\begin{center}
{\Large\bf {\boldmath Landau Gauge QCD: Functional Methods versus
  Lattice Simulations}\footnote{Plenary talk given at the 
 $13^\mathrm{th}$  International Conference 
on {\it Selected Problems of Modern Theoretical Physics} (SPMTP08),
Bogoliubov Laboratory of Theoretical Physics, Dubna, Russia, 23 -- 27
  June, 2008.}} 

\vspace*{6mm}

{Lorenz von Smekal}\\
{\small\it Centre for the Subatomic Structure of Matter, School of
  Chemistry and Physics,\\
 The University of Adelaide, SA 5005, Australia}
\end{center}

\vspace*{4mm}

\begin{abstract}
The infrared behaviour of QCD Green's functions in Landau gauge has
been focus of intense study. Different non-perturbative approaches 
lead to a prediction in line with the conditions for
confinement in local quantum field theory as spelled out in the 
Kugo-Ojima criterion. Detailed comparisons with lattice studies have
revealed small but significant differences, however. But aren't we
comparing apples with oranges when contrasting lattice Landau gauge
simulations with these continuum results? The answer is yes, and we need to
change that. We therefore propose a reformulation of Landau gauge on
the lattice which will allow us to perform gauge-fixed Monte-Carlo
simulations matching the continuum methods of local field theory which
will thereby be elevated to a truly non-perturbative level at the same
time.  
\end{abstract}

\vspace*{2mm}

\section*{\large Introduction} 

The Green's functions of QCD are the fundamental building blocks of
hadron phenomenology \cite{Alkofer:2000wg}.  Their infrared behaviour
is also known to contain essential information about the realisation
of confinement in the covariant formulation of QCD, in terms of local
quark and gluon field systems.  The Landau gauge Dyson-Schwinger
equation (DSE) studies of Refs.~\cite{vonSmekal:1997is,vonSmekal:1997vx}
established that the gluon propagator alone does not provide
long-range interactions of a strength sufficient to confine quarks.
This dismissed a widespread conjecture from the 1970's
going back to the work of Marciano, Pagels, Mandelstam and others. The
idea was revisited that the infrared dominant correlations are instead
mediated by the Faddeev-Popov ghosts of this formulation, whose
propagator was found to be infrared enhanced. This infrared behaviour
is now completely understood in terms of confinement in QCD
\cite{Alkofer:2000wg,Alkofer:2000mz,Lerche:2002ep}, it is a
consequence of the celebrated Kugo-Ojima (KO) confinement criterion. 

This criterion is based on the realization of the unfixed global gauge
symmetries of the covariant continuum formulation. In short, two
conditions are required by the KO criterion to distinguish confinement
from Coulomb and Higgs phases: (a) The massless single particle
singularity in the transverse gluon correlations of perturbation
theory must be screened non-perturbatively to avoid long-range fields
and charged superselection sectors as in QED. (b) The global gauge
charges must remain well-defined and unbroken to avoid the Higgs
mechanism. In Landau gauge, in which the (Euclidean) 
gluon and ghost propagators, 
\begin{equation}
     D^{ab}_{\mu\nu}(p) \,=\, 
  \delta^{ab}\left(\delta_{\mu\nu}-\frac{p_{\mu}p_{\nu}}{p^2}\right) 
  \frac{Z(p^2)}{p^2} \; , \;\;\; \mbox{and} \;\;
 D_G^{ab}(p)   \,=\,  -\delta^{ab}\;\frac{G(p^2)}{p^2} \; ,
\end{equation}
are parametrised by the two invariant functions $Z$ and $G$,
respectively, this criterion requires
\begin{equation}
\mbox{(a):} \;\; \lim_{p^2\to 0} Z(p^2)/p^2 \, < \,\infty \; ; \; \;\;
\mbox{(b):} \;\; \lim_{p^2\to 0} G^{-1}(p^2) \, =  \, 0 \; . 
\label{KO-IR}
\end{equation}
The translation of (b) into the infrared enhancement of the ghost
propagator (2b) thereby rests on the ghost/anti-ghost symmetry of the
Landau gauge or the symmetric Curci-Ferrari gauges. In particular,
this equivalence does not hold in linear covariant gauges with
non-zero gauge parameter such as the Feynman gauge.  

As pointed out in \cite{Lerche:2002ep}, the infrared enhancement of
the ghost propagator (2b) represents an additional boundary condition
on DSE solutions which then lead to the prediction of a conformal
infrared behaviour for the gluonic correlations in Landau gauge QCD
consistent with the conditions for confinement in local quantum field
theory. In fact, this behaviour is directly tied to the validity and
applicability of the framework of local quantum field theory for
non-Abelian gauge theories beyond perturbation theory. The subsequent
verification of this infrared behaviour with a variety of different
functional methods in the continuum meant a remarkable success. 
These methods which all lead to the same prediction 
include studies of their Dyson-Schwinger Equations (DSEs)
\cite{Lerche:2002ep}, Stochastic Quantisation \cite{Zwanziger:2001kw},
and of the Functional Renormalisation Group Equations
(FRGEs)~\cite{Pawlowski:2003hq}.  
This prediction amounts to infrared asymptotic forms
\begin{equation}
   \label{infrared-gh_gl}
  Z(p^2) \, \sim\, (p^2/\Lambda^2_{\mbox{\tiny QCD}})^{2\kappa_Z} \; , \;\; 
  \; \mbox{and} \;\;  G(p^2) \, \sim \, 
  (p^2/\Lambda_{\mbox{\tiny QCD}}^2)^{-\kappa_G} \; ,
\end{equation}
for $p^2 \to 0$, which are both determined by a unique critical
infrared exponent
\begin{equation}
  \label{kappaZ=kappaG}
   \kappa_Z = \kappa_G \equiv \kappa \; , 
\end{equation}
with $ 0.5 < \kappa < 1$. Under a mild regularity assumption
on the ghost-gluon vertex \cite{Lerche:2002ep}, the value of this
exponent is furthermore obtained as \cite{Lerche:2002ep,Zwanziger:2001kw} 
\begin{equation} 
\kappa \, = \, (93 - \sqrt{1201})/98 \, \approx \, 0.595 \; .
\label{kappa_c}
\end{equation}
The conformal nature of this infrared behaviour in the pure Yang-Mills
sector of Landau gauge QCD is evident in the generalisation to
arbitrary gluonic correlations \cite{Alkofer:2004it}: a uniform
infrared limit of one-particle irreducible vertex functions
$\Gamma^{m,n}$ with $m$ external gluon legs and $n$ pairs of
ghost/anti-ghost legs of the form
\begin{equation}
\Gamma^{m,n} \, \sim\, (p^2/\Lambda_{\mbox{\tiny QCD}}^2)^{(n-m)\kappa}\; ,   
\label{genIR}
\end{equation}
when all $p_i^2 \propto p^2 \to 0$, $i=1,\dots 2n+m$. In particular,
the ghost-gluon vertex is then infrared finite (with $n=m=1$) as it
must \cite{Taylor:1971ff}, and the non-perturbative running coupling 
introduced in \cite{vonSmekal:1997is,vonSmekal:1997vx} via the
definition
\begin{equation} 
 \alpha_S(p^2) \, = \, \frac{g^2}{4\pi} Z(p^2) G^2(p^2) 
\label{alpha_minimom}
\end{equation}
approaches an infrared fixed-point, $\alpha_S \to \alpha_c$ for $p^2 \to
0$. If the ghost-gluon vertex is regular at $p^2 =0$, its value is
maximised and given by \cite{Lerche:2002ep}
\begin{equation} 
\alpha_c \, = \, \frac{8\pi}{N_c} \, \frac{\Gamma^2(\kappa-1)
  \Gamma(4-2\kappa)}{\Gamma^2(-\kappa) \Gamma(2\kappa-1)} \,  \approx
  \, \frac{9}{N_c} \times \, 0.99  \; .  
\label{alpha_c}
\end{equation}
Comparing the infrared scaling behaviour of DSE and FRGE solutions of
the form of Eqs.~(\ref{infrared-gh_gl}), it has in fact been shown
that in presence of a single scale, the QCD scale
$\Lambda_{\mbox{\tiny QCD}}$, the solution with the infrared behaviour
(\ref{kappaZ=kappaG}) and (\ref{genIR}), with a positive exponent
$\kappa$,  is unique \cite{Fischer:2006vf}. Because of its 
uniqueness, it is nowadays being called the {\em scaling solution}. 

This uniqueness proof does not rule out, however, the possibility of a
solution with an infrared finite gluon propagator, as arising from a
transverse gluon mass $M$, which then leads to an essentially free ghost
propagator, with the free massless-particle singularity at $p^2=0$,
{\it i.e.},
\begin{equation} 
Z(p^2) \, \sim \, p^2/M^2\; , \;\; \mbox{and} \;\; G(p^2) \, \sim \,
\mathrm{const.}\; \label{decoupling}
\end{equation}
for $p^2 \to 0$. The constant contribution to the zero-mo\-men\-tum gluon
propagator, $ D(0) = 3/(4 M^2)$, thereby necessarily leads to an
infrared constant ghost renormalisation function $G$. This
solution corresponds to $\kappa_Z = 1/2 $ and $ \kappa_G = 0$. It does
not satisfy the scaling relations (\ref{kappaZ=kappaG}) or (\ref{genIR}).
This is because in this case the transverse gluons decouple
for momenta $p^2 \ll M^2 $, below the independent second scale given
by their mass $M$. It is thus not within the class of scaling
solutions considered above, and it is termed the {\em decoupling
  solution} in contradistinction \cite{Fischer:2008uz}. The
interpretation of the renormalisation group invariant
(\ref{alpha_minimom}) as a running coupling does not make sense in the
infrared in this case, in which there is no infrared fixed-point and
no conformal infrared behaviour.   

Without infrared enhancement of the ghosts in Landau gauge, the global
gauge charges of covariant gauge theory are spontaneously
broken. Within the language of local quantum field theory the
decoupling solution can thus only be realised if and only if it 
comes along with a Higgs mechanism and massive physical gauge bosons. 
The Schwinger mechanism can in fact be described in this way, and it
can furthermore be shown that a non-vanishing gauge-boson mass, by whatever
mechanism it is generated, necessarily implies the spontaneous
breakdown of global symmetries \cite{Nakanishi:1990}.

\section*{\large Landau Gauge QCD in the Continuum and on the
      Lattice}

Early lattice studies of the gluon and ghost propagators 
supported their predicted infrared behaviour qualitatively well.
Because of the inevitable finite-volume effects, however, these results 
could have been consistent with both, the scaling solution as well as  the
decoupling solution. Recently, the finite-volume effects have been
analysed carefully in the Dyson-Schwinger equations to demonstrate how
the scaling solution is approached in the infinite volume limit there
\cite{Fischer:2007pf}. Comparing these finite volume DSE results with 
latest $SU(2)$ lattice data on impressively large lattices
\cite{Sternbeck:2007ug,Cucchieri:2007md}, corresponding to physical
lengths of up to $20$~fm in each direction, finite-volume effects
appear to be ruled out as the dominant cause of the observed
discrepancies with the scaling solution. The lattice results are much
more consistent with the decoupling solution which poses the obvious
question whether there is something wrong with our general
understanding of covariant gauge theory or whether we are perhaps 
comparing apples with oranges when applying inferences drawn from
the infrared behaviour of the lattice Landau gauge correlations 
on local quantum field theory? 

The latter language is based on a cohomology construction of a
physical Hilbert space over the indefinite metric spaces of covariant
gauge theory from the representations of the Becchi-Rouet-Stora-Tyutin
(BRST) symmetry. But do we have a non-perturbative definition of a
BRST charge? The obstacle is the existence of the so-called Gribov copies
which satisfy the same gauge-fixing condition, {\it i.e.}, the Lorenz
condition in Landau gauge, but are related by gauge transformations,
and are thus physically equivalent. In fact, in the direct translation
of BRST symmetry on the lattice, there is a perfect cancellation among
these gauge copies which gives rise to the famous Neuberger $0/0$
problem. It asserts that the expectation value of any gauge invariant
(and thus physical) observable in a lattice BRST formulation will
always be of the indefinite form $0/0$ \cite{Neuberger1987} 
and therefore prevented such formulations for more than 20 years now. 

In present lattice implementations of the Landau gauge this problem is
avoided because the numerical procedures are based on minimisations of
a gauge fixing potential w.r.t.~gauge transformations. To find
absolute minima is not feasable on large lattices as this is a
non-polynomially hard computational problem. One therefore settles for
local minima which in one way or another, depending on the algorithm,
samples gauge copies of the first Gribov region among which there is
no cancellation. For the same reason, however, this is not a BRST
formulation. The emergence of the decoupling solution can thus not be
used to dismiss the KO criterion of covariant gauge theory in the continuum.

\section*{\large Strong Coupling Limit of Lattice Landau Gauge}

From the finite-volume DSE solutions of \cite{Fischer:2007pf} it
follows that a wide separation of scales is necessary before one can
even hope to observe the onset of an at least approximate conformal
behaviour of the correlation functions in a finite volume of
length $L$. What is needed is a reasonably large number of modes with
momenta $p$ sufficiently far below the QCD scale $\Lambda_{\mbox{\tiny
    QCD}}$ whose corresponding wavelengths are all at the same time
much shorter than the finite size $L$,
\begin{equation}
               \pi/L \, \ll p \, \ll \Lambda_{\mbox{\tiny QCD}} \; .
\label{scales}
\end{equation}
It was estimated that this requires sizes $L$ of about $15$ fm,
especially for a power law of the ghost propagator of the form in 
(\ref{infrared-gh_gl}) to emerge in a momentum range with
(\ref{scales}). A reliable quantitative determination of the exponents
and a verification of their scaling relation (\ref{kappaZ=kappaG}) on
the other hand might even require up to $L = 40$ fm \cite{Fischer:2007pf}.
 
As an alternative to the brute-force method of using ever larger lattice
sizes for the simulations might therefore be to ask what one observes
when the formal limit $\Lambda_{\mbox{\tiny QCD}} \to \infty $ is
implemented by hand. This should then allow to assess whether the
predicted conformal behaviour can be seen for the larger lattice
momenta $p$, after the upper bound in (\ref{scales}) has been removed,
in a range where the dynamics due to the gauge action would otherwise
dominate and cover it up completely. Therefore, the ghost and gluon
propagators of pure $SU(2)$ lattice Landau gauge were studied in the
strong coupling limit $\beta \to 0$ in \cite{SternbeckIP,Sternbeck:2008wg}. 

\begin{figure}[t]
\leftline{\includegraphics[width=0.5\linewidth]{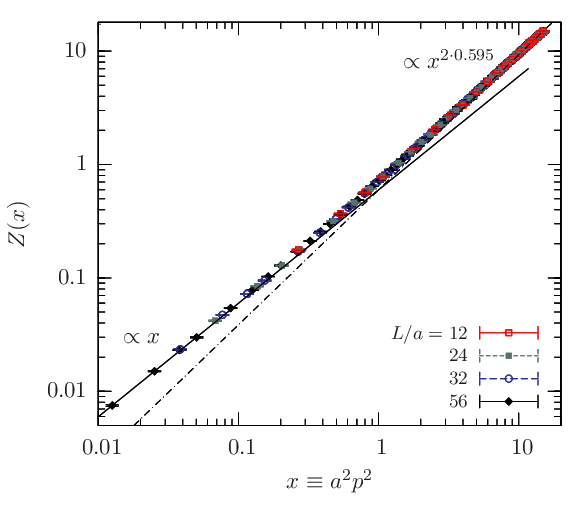}
\hskip .2cm
\includegraphics[width=0.5\linewidth]{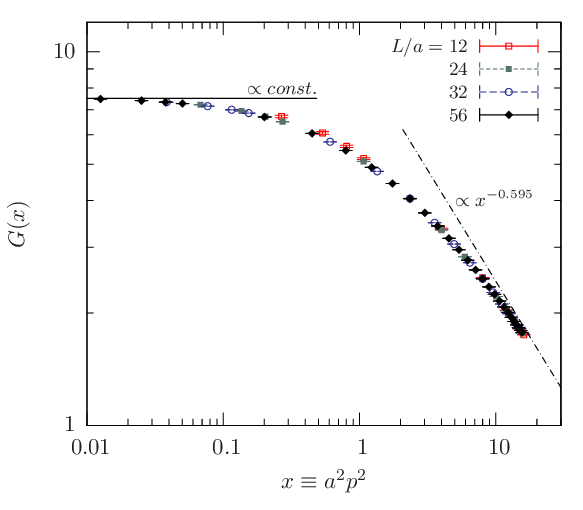}
}
\caption{The gluon (left) and ghost (right) dressing functions
  at $\beta = 0$ compared to the decoupling solution 
  (solid) and the scaling solution 
  (dashed) with $\kappa $
  from (\ref{kappa_c}) (not fitted).}
\label{beta0props}
\end{figure}

In this limit, the gluon and ghost dressing functions
tend towards the decoupling solution at small momenta and towards the
scaling solution at large momenta (in units of the lattice spacing
$a$) as seen in Figure \ref{beta0props}. The transition
from decoupling to scaling occurs at around $a^2 p^2 \approx 1$,
independent of the size of the lattice. The observed deviation from
scaling at $ a^2p^2 < 1$ is thus not a finite-size effect. The high
momentum branch can be used to attempt fits of $\kappa_Z$ and
$\kappa_G$ in (\ref{infrared-gh_gl}) and the data is consistent with
the scaling relation (\ref{kappaZ=kappaG}). With some dependence on the
model used to fit the data, good global fits are generally obtained for 
$\kappa = 0.57(3)$, with very little dependence on the lattice size. 

For the scaling solution one would expect the running coupling 
defined by (\ref{alpha_minimom}) to approach its constant fixed-point
value in the strong-coupling limit, and this is indeed being observed 
for the scaling branch \cite{SternbeckIP}: The numerical data for the product
(\ref{alpha_minimom}) levels at $\alpha_c \approx 4$ for large $a^2
p^2$. As expected for an exponent $\kappa$ slightly smaller than the
value in (\ref{kappa_c}), see \cite{Lerche:2002ep}, this is just below the
upper bound given by (\ref{alpha_c}), $\alpha_c \approx 4.45$ for $SU(2)$. 

When comparing various definitions of gauge fields on the lattice,
all equivalent in the continuum limit, one furthermore observes that
neither the estimate of the critical exponent $\kappa $ nor the
corresponding value of $\alpha_c$ are sensitive to the definition
used \cite{SternbeckIP}. This is in contrast to the decoupling branch
for $a^2 p^2 < 1$, which is very sensitive to that
definition. Different definitions, at order $a^2$ and beyond, lead to
different Jacobian factors. This is well known from lattice
perturbation theory where, however, the lattice Slavnov-Taylor
identities guarantee that the gluon remains massless at every order
by cancellation of all quadratically divergent contributions to its
self-energy. The strong-coupling limit, where the effective mass in
(\ref{decoupling}) behaves as $M^2 \propto 1/a^2$, therefore shows
that such a contribution survives non-perturbatively in minimal
lattice Landau gauge. This contribution furthermore depends on the
measure for gauge fields whose definition from minimal lattice Landau
is therefore ambiguous. One might still hope that this ambiguity will
go away at non-zero $\beta$, in the scaling limit. While this is true
at large momenta, it is not the case in the infrared, at least not for 
commonly used values of the lattice coupling such as
 $\beta = 2.5$ or $\beta = 2.3$ in $SU(2)$, as demonstrated in
\cite{SternbeckIP}.

\section*{\large \boldmath Lattice BRST and the Neuberger $0/0$
  Problem}

It would obviously be desirable to have a BRST symmetry on the lattice which
could then provide lattice Slavnov-Taylor identities beyond
perturbation theory. In principle, this could be achieved by inserting
the partition function of a topological model with
BRST exact action into the gauge invariant lattice measure. Because of
its topological nature, this gauge-fixing partition function
$Z_{\mbox{\tiny GF}}$ will be independent of gauge orbit and gauge
parameter. The problem is that in the standard formulation this
partition function calculates the Euler characteristic $\chi $ 
of the lattice gauge group which vanishes \cite{Schaden1998},
\begin{equation}
Z_{\mbox{\tiny GF}} = \chi(SU(N)^{\#{\rm sites}}) =
\chi(SU(N))^{\#{\rm sites}} = 0^{\# {\rm sites}}\; . 
\label{EulerC}
\end{equation}
Neuberger's $0/0$ problem of lattice BRST arises because we have then
inserted zero instead of unity (according to the Faddeev-Popov
prescription) into the measure of lattice gauge theory.
On a finite lattice, such a topological model is equivalent to a
problem of supersymmetric quantum mechanics with Witten index 
$\mathcal{W} = Z_{\mbox{\tiny GF}}$. 
Unlike the case of primary interest in supersymmetric quantum
mechanics, here we need a model with non-vanishing Witten
index to avoid the Neuberger $0/0$ problem. Then however, just as the
supersymmetry of the corresponding quantum mechanical model, such a
lattice BRST cannot break.

In Landau gauge, with gauge parameter $\xi = 0$, the Neuberger zero,
$Z_{\mbox{\tiny GF}} = 0$, arises from the perfect cancellation of Gribov
copies via the Poincar\'e-Hopf theorem. The gauge-fixing potential
$V_U[g]$ for a generic link configuration $\{ U\}$  
thereby plays the role of a Morse potential for gauge transformations $g$
and the Gribov copies are its critical points (the global gauge
transformations need to remain unfixed so that there are
strictly speaking only $(\# $sites$-1)$ factors of $\chi(SU(N)) =
0$ in (\ref{EulerC})). The Morse inequalities
then immediately imply that there are at least $2^{(N-1)(\# {\rm
    sites}-1)}$ such copies in $SU(N)$ on the lattice, or $2^{\# {\rm
    sites}-1}$ in compact $U(1)$, and equally many with either sign
of the
Faddeev-Popov determinant ({\it i.e.}, that of the Hessian of $V_U[g]$).  

The topological origin of the zero originally observed by Neuberger in a
certain parameter limit due to uncompensated Grassmann ghost
integrations in standard Faddeev-Popov theory \cite{Neuberger1987} 
becomes particularly
evident in the ghost/anti-ghost symmetric Curci-Ferrari gauge with its
quartic ghost self-interactions \cite{Sme08}. Due to its Riemannian
geometry with symmetric connection and curvature tensor $R_{ijkl} =
\frac{1}{4} {f^a}_{ij} {f^a}_{kl}$ for $SU(N)$, in this gauge the same
parameter limit leads to computing the zero in (\ref{EulerC}) from a
product of independent Gauss-Bonnet integral expressions,
\begin{equation}
   \chi\big(SU(N)\big) \, = \, 
   \frac{1}{(2\pi)^{(N^2\!-1)/2}\hskip -1.3cm}\hskip 1.4cm 
  \int_{SU(N)} \hskip -.2cm dg \, \int d\bar c\, dc \; \exp\Big\{  
\smallfrac{1}{4}\, R_{abcd} \, \bar c^a\bar c^b  c^c c^d \Big\} \, =
  \, 0 \; ,    \label{GaussBonnet}
\end{equation}
for each site of the lattice. This corresponds to the Gauss-Bonnet
limit of the equivalent supersymmetric quantum mechanics model in
which only constant paths contribute \cite{Birmingham1991}.

The indeterminate form of physical observables
as a consequence of (\ref{GaussBonnet}) is regulated by a
Curci-Ferrari mass term. While such a mass $m$ decontracts the double
BRST/anti-BRST algebra, which is well-known to result in a loss of unitarity,
observables can then be meaningfully defined in the limit  
$m \to 0$ via l'Hospital's rule \cite{Sme08}.

\section*{\large Lattice Landau Gauge from Stereographic Projection}

The 0/0 problem due to the vanishing Euler characteristic of $SU(N)$  
is avoided when fixing the gauge only up to the maximal Abelian
subgroup $U(1)^{N-1}$ because the Euler characteristic of the coset
manifold is non-zero. The corresponding lattice BRST has been
explicitly constructed for $SU(2)$ \cite{Schaden1998}, where the coset
manifold is the 2-sphere and $\chi(SU(2)/U(1)) = \chi(S^2) = 2$. 
This indicates that the Neuberger problem might be solved when
that of compact $U(1)$ is, where the same cancellation of lattice Gribov
copies arises because $\chi(S^1) = 0$. A surprisingly simple
solution to this problem is possible, however, by stereographically
projecting the circle $S^1 \to \mathbb{R}$ which can be achieved by a
simple modification of the minimising potential
\cite{vonSmekal:2007ns}.  The resulting potential is convex to the
above and leads to a positive definite Faddeev-Popov operator for
compact $U(1)$ where there is thus no cancellation of Gribov
copies, but $Z_{\mbox{\tiny GF}}^{U(1)} = N_{\mbox{\tiny GC}}$, for
$N_{\mbox{\tiny GC}}$ Gribov copies. 

As compared to the standard lattice Landau gauge for compact
$U(1)$ their number is furthermore exponentially reduced. This is easily
verified explicitly in low dimensional models.
While $N_{\mbox{\tiny GC}}$ grows exponentially with the number of sites in
the standard case as expected, the stereographically projected version
has only $N_{\mbox{\tiny GC}} = N_x $ copies on a periodic chain of length
$N_x$  and $\ln N_{\mbox{\tiny GC}} \sim N_t \ln N_x $ on a 
$2D$ lattice of size $N_t   N_x$ in Coulomb gauge, for example, 
and in both cases
their number is verified to be independent of the gauge orbit. 
 
The general proof of  $Z_{\mbox{\tiny GF}}^{U(1)} = N_{\mbox{\tiny GC}}$ with
stereographic projection which avoids the Neuberger zero in
compact $U(1)$ \cite{vonSmekal:2007ns} follows from a simple example
of a Nicolai map \cite{Birmingham1991}.   

Applying the same techniques to the maximal Abelian subgroup
$U(1)^{N-1}$, the generalisation to $SU(N) $ lattice gauge theories is
possible when the odd-dimensional spheres $S^{2n+1} $, $n=1,\dots
N\!-\!1$, of its parameter space are stereographically projected to
$\mathbb{R} \times \mathbb{R}P(2n)$. In absence of the cancellation of
the lattice artifact Gribov copies along the $U(1)$ circles, the
remaining cancel\-lations between copies of either sign in $SU(N)$,
which will persist in the continuum limit, are then necessarily
incomplete, however, because $\chi(\mathbb{R}P(2n)) = 1$. 

For $SU(2)$ this program is straightforward. One replaces the standard
 gauge-fixing potential $V_U[g]$ of lattice Landau gauge by
 $\widetilde V_U[g]$, via gauge-transformed links $U^g_{x\mu}$, where
\begin{equation}
   V_U[g] = 4 \sum_{x,\mu}\left(1-\smallfrac{1}{2} \Tr
    U^g_{x\mu}\right)
 \quad \mbox{and} \quad   
  \label{modLG}
  \widetilde V_U[g] = -8 \sum_{x,\mu}
  \ln\left(\frac{1}{2}+\frac{1}{4}\Tr 
    U^{g}_{x,\hat{\mu}} \right)\;. 
\end{equation}
The standard and stereographically projected
gauge fields on the lattice are defined as
\[
   {A}_{x\mu} = \frac{1}{2ia}\left({U}_{x\mu} -
    {U}^{\dagger}_{x\mu}\right)  \quad \mbox{and} \quad 
\widetilde{A}_{x\mu} = \frac{1}{2ia}\left(\widetilde{U}_{x\mu} -
    \widetilde{U}^{\dagger}_{x\mu}\right)
  \; , \;\; \text{with}\quad
  \widetilde{U}_{x\mu} 
     \equiv \frac{2 U_{x\mu}}{1+ \frac{1}{2} \Tr U_{x\mu}} \; . 
  \label{gauge-fields}
\]
The gauge-fixing conditions $F = 0$ and $\widetilde F = 0$ 
are their respective lattice divergences, in the language of lattice
cohomology, $F = \delta A$ and $\widetilde F = \delta \widetilde A$.
A particular advantage of the non-compact $\widetilde A$ is that they
allow to resolve the modified lattice Landau gauge condition
$\widetilde F = 0$ by Hodge decomposition. This provides a framework
for gauge-fixed Monte-Carlo simulations which is currently being
developed for the particularly simple case of $SU(2)$ in 2 dimensions.
In the low-dimensional models mentioned above it can
furthermore be verified explicitly that the corresponding topological
gauge-fixing partition function is indeed given by 

\vspace{-.7cm}
\begin{equation}
Z_{\mbox{\tiny GF}}^{SU(2)} 
   \, = \,  Z_{\mbox{\tiny GF}}^{U(1)}  \, \not= \, 0   \; , 
\label{ModZgf}
\end{equation}
as expected from $\chi(\mathbb{R}P(2)) = 1$. The proof of this will be
given elsewhere.

\section*{\large Conclusions and Outlook}

Comparisons of the infrared behaviour of QCD Green's functions as
obtained from lattice Landau gauge implementations based on
minimisations of a gauge-fixing potential and from continuum studies
based on BRST symmetry have to be taken with a grain of salt.  
Evidence of the asymptotic conformal behaviour predicted by the latter
is seen in the strong coupling limit of lattice Landau gauge where
such a behaviour can be observed at large lattice momenta $a^2p^2\gg
1$. There the strong coupling data is consistent with the predicted
critical exponent and coupling from the functional approaches. The
deviations from scaling at $a^2 p^2 < 1$ are not finite-volume effects,
but discretisation dependent and hint at a breakdown of BRST symmetry
arguments beyond perturbation theory in this
approach. Non-perturbative lattice BRST has been plagued by the
Neuberger $0/0$ problem, but its improved topological understanding provides
ways to overcome this problem. The most promising one at this point
rests on stereographic projection to define gauge fields on the
lattice together with a modified lattice Landau gauge. This new
definition has the appealing feature that it will allow
gauge-fixed Monte-Carlo simulations in close analogy to the continuum
BRST methods which it will thereby elevate to a non-perturbative
level.


\end{document}